\documentclass[letterpaper,twocolumn,10pt]{article}
\usepackage{usenix-2020-09}

\usepackage[square,numbers,sort&compress]{natbib}
\usepackage{booktabs}
\usepackage{multirow}
\usepackage[utf8]{inputenc}
\usepackage{amssymb, amsmath}
\usepackage{ulem}
\usepackage{tikz}
\usepackage{comment}
\usepackage{xspace}
\usepackage{multirow} 
\usepackage{hyperref}
\usepackage{subcaption}
\usepackage{enumitem}

\usepackage{footnote}
\usepackage{csquotes}
\usepackage{utfsym}

\usepackage{appendix}[title]

\renewcommand{\sectionautorefname}{\S\kern-0.2em}
\renewcommand{\subsectionautorefname}{\S\kern-0.2em}
\renewcommand{\subsubsectionautorefname}{\S\kern-0.2em}

\iffalse

	\newcommand{\omer}[1]{\textcolor{red}{#1}}
	\newcommand{\richie}[1]{\textcolor{red}{#1}}
	\newcommand{\michelle}[1]{\textcolor{teal}{MLM: #1}}

	\newcommand{\dml}[1]{\textcolor{magenta}{\textbf{DL: #1}}}
	\newcommand{\todo}[1]{\textcolor{red}{\textbf{TODO: #1}}}
\else
	\newcommand{\michelle}[1]{}
	\newcommand{\omer}[1]{}
	\newcommand{\dml}[1]{\textcolor{magenta}{\textbf{I WILL NOT BE SILENCED}}}
	\newcommand{\richie}[1]{}
	\renewcommand{\dml}[1]{\textcolor{magenta}{\textbf{Ok fine}}}
	\renewcommand{\dml}[1]{}
	\newcommand{\todo}[1]{}
\fi

\iffalse
	\newcommand{\new}[1]{\textcolor{red}{#1}}
	\newcommand{\note}[1]{\textcolor{blue}{#1}}
	\newcommand{\x}[1]{\sout{\textcolor{red}{#1}}}
\else
	\newcommand{\new}[1]{#1}
	\newcommand{\note}[1]{#1}
	\newcommand{\x}[1]{}
\fi

\renewcommand{\emph}[1]{\textit{#1}}

\renewcommand{\sp}{security \& privacy\xspace}

\newcommand{\Sp}{Security \& privacy\xspace}

\newcommand{\authorand}{\end{tabular}\hspace{.5em}\begin{tabular}[t]{c}}

\newcommand{\yt}{YouTube\xspace}

\newcommand{\yters}{YouTubers\xspace}

\newcommand{\vpnads}{VPN ads\xspace}
\newcommand{\vpnad}{VPN ad\xspace}

\newcommand{\factualshort}{factual VPN\xspace}
\newcommand{\misleadingshort}{misleading VPN\xspace}
\newcommand{\allshort}{all VPN\xspace}
\newcommand{\threatshort}{threat\xspace}

\newcommand{\misleadingmm}{\misleadingshort mental models\xspace}
\newcommand{\allmm}{\allshort models\xspace}
\newcommand{\threatmm}{\threatshort mental models\xspace}

\let\oldsection\subsubsection
\renewcommand{\subsubsection}[1]{\vspace{4mm} \oldsection{#1}  }

\renewcommand{\paragraph}[1]{ \medskip\noindent\textbf{#1\phantom{xxx}}}

\urldef{\ytarticleEU}\url{https://support.google.com/youtube/thread/17592587/%F0%9F%9A%A9-updates-on-article-17-formerly-article-13?hl=en}

\urldef{\googletrendurl}\url{https://trends.google.com/trends/explore?date=2019-10-01%202020-09-30&geo=US&q=%2Fm%2F012t0g}

\begin{document}

\title{As Advertised? Understanding the Impact of Influencer VPN Ads} 

\author{
{\rm Omer Akgul}\textsuperscript{$\diamond$}\textsuperscript{$\usym{1F319}$}
\authorand
{\rm Richard Roberts}\textsuperscript{$\diamond$}
\authorand
{\rm Emma Shroyer}\textsuperscript{$\diamond$}
\authorand
{\rm Dave Levin}\textsuperscript{$\diamond$}
\authorand
{\rm Michelle L. Mazurek}\textsuperscript{$\diamond$}
\and
\textsuperscript{$\diamond$}\textit{University of Maryland}\hspace{1cm}\textsuperscript{$\usym{1F319}$}\textit{Carnegie Mellon University}\hspace{1cm}
}

\maketitle

\begin{abstract}
Influencer VPN ads (sponsored segments)
on YouTube often disseminate misleading information about both VPNs,
and security \& privacy more broadly.
However, it remains unclear how (or whether) these ads
affect users' perceptions and knowledge about VPNs.
In this work, we explore the relationship between YouTube VPN ad exposure
and users' mental models of VPNs, security, and privacy.
We use a novel VPN ad detection model to calculate the ad exposure of
217 participants via their YouTube watch histories, and we develop
scales to characterize their mental models in relation to claims commonly
made in VPN ads.
Through (pre-registered) regression-based analysis, we find that 
exposure to VPN ads is significantly correlated with 
familiarity with VPN brands and increased belief in 
\note{(hyperbolic)} threats. 
While not specific to VPNs, these threats are often discussed in VPN ads.
In contrast, 
\new{although many participants agree with both factual and 
misleading mental models of VPNs that often appear in ads, 
we find no significant correlation between exposure to VPN 
ads and these mental models.}
\new{These findings suggest that, if VPN ads do impact mental models, 
then it is predominantly emotional 
(i.e., threat perceptions) rather than technical.}

\end{abstract}

\section{Introduction}
\label{sec:intro}

Do ads actually change the way people think?
This simple question has been central to a vast body of academic
literature in the fields of marketing and advertising, leading to
myriad findings about how the frequency and messaging of ads can impact
a viewer's brand awareness and willingness to purchase a given product 
(e.g.,~\cite{lee2010exploring, schmidt2015advertising, 
chatterjee2003modeling, cox2002beyond, deighton1994radical, 
hassan2021logtruth, manchanda2006banner, simon1982adpuls, 
betts2019taking, yaveroglu2008advertising, chae2019wearout}).
Similarly, public health experts have studied the impact that ads can
have on health choices, such as how prescription drug ads influence 
patients' requests for the advertised drugs~\cite{defrank2020direct}.

Recently this question has also become critical to the
security community, with the proliferation of advertising for security 
and privacy tools. Influencer VPN ads\footnote{Unlike interstitial ads 
(delivered by \yt \new{before, during, and after videos}), influencer ads are part of the video, 
typically produced by influencers\new{, integrated into the content, and without an explicit business tie to \yt}.}
on \yt, in particular, have recently 
become near-ubiquitous. 
These ads tend to include a significant amount of
``educational'' content, informing viewers of potential attacks and the
defenses that VPNs provide.
We had previously studied the content of a random sample of influencer VPN ads,
finding extensive instances of misleading or false claims~\cite{akgul2022vpn}, such as that using a
VPN means ``you won't ever have to worry about anything on the internet
again''~\cite{yt2020dontworrybehappy}.
Unfortunately, this misinformation is not limited to a small audience; 
we previously 
estimated that influencer VPN ads have received over \emph{4.5
billion views} on \yt alone.
This previous study
takes a first step towards understanding these ads, 
but provides no evidence on user impact. Is 
our 
speculation 
on internet threat and VPN mental model influence correct?

In this paper, we look beyond the basic question of whether VPN ads
influence viewers to purchase products and ask: what
impact do they have on viewers' \sp mental models?
To the best of our knowledge, we are the first to seek to directly
measure the impact of \emph{any} entertainment
media on viewers' \sp mental models.
Prior studies \new{investigating} mental models~\cite{fulton2019media,
redmiles2016learn, baig2021media, tahaei2021don, redmiles2016think,
namara2020emotional} often report media depictions as a top
source of \sp information~\cite{redmiles2016learn, routi2017posture}.
However, these findings are based on users' self-reported data, 
which, while valuable, is subject to limitations in what participants
can (and are willing to) recall.

We introduce a novel, multi-stage user study design in which we
gather not only participants' mental models, but also \emph{their
entire \yt viewing history}.
In total, we collected this data
from 217 \yt users.
Leveraging this rich data, we expand on self-reported data by 
\emph{directly measuring} real-world ad
exposure and investigate its correlation with \sp mental models.

We find significant correlations between exposure to \vpnads and both
VPN brand familiarity and belief in \note{(hyperbolic)} threats. 
However, we find no such relationship between exposure to \vpnads and
belief in specific factual or misleading claims about VPN capabilities \new{that commonly appear in \vpnads}.
Qualitative analysis of our data suggests that participants hold similar, but not
identical, mental models to those disseminated through \vpnads.

\new{These findings suggest two possible (not mutually exclusive) 
interpretations: that VPN ads have an impact on influencing 
users threat models, or that advertisers (VPN companies, the \yters, or 
the \yt content delivery algorithms) intentionally invoke 
threats that resonate with and thereby reinforce models the 
viewers already have. In either case, the results indicate 
that emotion plays a larger role than the technical 
details in VPN ads~\cite{akgul2022vpn, obermiller2005ad}.}

\paragraph{Contributions} 
We make the following contributions:
\begin{itemize}[itemsep=0pt, left=0pt]
\item We introduce a novel user study design to collect \vpnad-relevant
mental models and measure users' \vpnad exposure using entire \yt viewing histories.
We collect a dataset of 217 \yt users using this method.
\item We demonstrate, for the first time, real-world (not merely
self-reported) \new{correlation between} entertainment media \new{and} 
users' security mental models.
\item We show that \yt users are extensively exposed to
\vpnads. This exposure strongly correlates with brand
familiarity and belief in \note{hyperbolic}
threats advertised.
\end{itemize}

\section{Background and Related Work}
\label{sec:related}
In this section we summarize related work on VPN mental models and use
cases, the use of \yt as an information source for users, and provide
background on modeling the influence of advertisements.

\paragraph{VPN uses and mental models} 
User mental models about VPNs have received attention from usable security
researchers, motivated by increasing numbers of commercial VPNs and
their consumers. 
\new{Mental models are typically defined as users' internal 
representations of reality~\cite{jones2011mental}. 
We use a more constrained definition, common in \sp work: 
a user's comprehension of how a 
system operates, including its inputs, outputs, and 
the expected effects.} %
Namara et al.\ found that tech-savvy VPN users tended to use 
VPNs for non-privacy-related use cases like accessing 
geo-locked content~\cite{namara2020emotional}. 
However, the same study found that users motivated by privacy tend to use
VPNs for \textit{longer}.
Ramesh et al.~\cite{ramesh2023vpn} found contradictory results investigating
another large (likely) tech-savvy population: the majority of their 
participants reported using VPNs for security reasons.
They also find that users rely on recommendation websites when choosing
\textit{which} VPN to use, but it remains unclear how users \textit{initially} learn about VPNs.
Dutkowska-Zuk et al.~find greater concern with content access and less
concern with \sp, with students compared to 
a crowd-sourced general population~\cite{dutkowska2022vpn}. 
Again contradicting prior work, they find that censorship avoidance is
a major motivation for U.S. users.

Story et al.~found misaligned VPN mental models to be common in a 
demographically representative online sample~\cite{story2021awareness}.
Researchers have noted users adopting VPNs to increase \sp while on public
networks, as well as to prevent hacks or password 
leaks~\cite{zou2020examining}.
\new{Researchers have found that roughly 
6\% of Tor users use VPNs in conjunction, potentially 
indicating the influence of popular media~\cite{fassl2023vpnfolklore}.
While using VPNs to access Tor might provide benefits under narrow 
threat models (e.g., preferring the ISP to know a VPN 
connection is happening instead of Tor, the VPN is trustable), in many cases 
it provides no benefits and, depending on the trustworthiness 
of VPN~\cite{ikram2016androidvpn, hotspot2017ftccomplaint, ramesh2022vpnalyzer}, 
may even harm privacy~\cite{fassl2023vpnfolklore}.}
Our prior work reported \yters conveying similar (dubious) ideas about VPN
features and benefits~\cite{akgul2022vpn}, hinting at the possible influence
of \vpnads on mental models.
Note that VPNs are not a panacea for security and privacy;
prior work has found some VPNs that range from misconfigured, to dishonest and 
even malicious~\cite{ikram2016androidvpn, weinberg2018proxies, 
bischoff2020zerologs, ng2021vpnprivate, khan2018commercialvpn, 
hotspot2017ftccomplaint,ramesh2022vpnalyzer}.

Our work builds on this body of work by investigating the (potential)
\textit{source} of these misaligned mental models.
Some of the same themes from prior work are replicated in our data.

\paragraph{Information on \yt}
Researchers have found \yt to be the top online resource for users 
seeking information about multiple topics~\cite{kross2021characterizing}
including topics that users otherwise know nothing about~\cite{smith2018many}.
This makes \yt an interesting object of study for \sp education,
as prior security research has found that some users build their mental
models of \sp using (online) media~\cite{fulton2019media, baig2021media,
redmiles2016learn} and ads~\cite{routi2017posture}.

We had previously speculated about the impact \yt might have on mental models
through the context of \vpnads~\cite{akgul2022vpn}.
Through a random sample of \vpnads, we had found that these ads are a potential
source of \sp education that reaches a broad audience.
Troublingly, we had find that some influencer VPN ads contain vague, misleading,
and/or false information about both VPNs and internet threats, threatening
to have a harmful influence on viewers' mental models and security behaviors.
These findings may be amplified by the influencer effect: influencer ads
are both powerful and cost-effective ways to influence
consumers~\cite{lou2019investigating, nazerali2017ytinfluencers, 
yuan2020social, lou2019fancying}.

Motivated by these findings, our research aims to directly investigate the
impact of \vpnads, by looking for correlations between user mental models
and exposure to prominent advertising themes
we previously reported~\cite{akgul2022vpn}.

\paragraph{\new{Misleading ads}} \new{At a broader level, ads that are inappropriate, 
deceptive, or otherwise manipulative are commonplace on the internet. 
Researchers have uncovered such ads in multiple contexts: 
on social media targetting 
minorities~\cite{ali2023problematic, ali2019discrimination},
on news websites before elections~\cite{zeng2021polls}, potentially 
illegally on childrens websites~\cite{moti2024targeted}, and more.}

\paragraph{Risk Communication}
According to the human-in-the-loop framework, users act
as receivers of communications (e.g., warnings) which
may influence their behavior~\cite{cranor2008loop}.
An extensive body of literature has investigated different
ways to effectively convey risks and 
threats to users. Notably, communication efforts are often 
hindered by user habituation~\cite{bravolillo2014attractors}
Bravo-Lillo et al. demonstrated that attractors, or attention capturing UI elements, 
could reduce warning habituation~\cite{bravolillo2014attractors}.
In 2017, Albayram et al. showed that fear appeals were effective
at inducing behavior change~\cite{albayram2017fear}.
In a replication of that study, Qahtani et al. found that fear
appeals were especially effective when targeting the shared
fears of a specific population~\cite{qahtani2018fear2}.
Effective risk communication can even influence purchasing 
decisions; researchers showed that users were willing to 
pay a premium for privacy-preserving IoT products given appropriate 
risk communication~\cite{gopavaram2019iotmarketplace, emami2023consumers}.

We extend this body of work by investigating VPN ads as risk 
communications that frequently include fear appeals, are often targeted
to the audience of a specific \yt channel, and are inherently 
designed to persuade users to purchase a brand's product.
In contrast with prior work, we investigate the relationship between
natural exposure to VPN ads in the wild
and users' mental models of security threats.

\paragraph{Modeling the Influence of Ads}
\label{sec:modeling_ads}
Modeling the effects of repeated ad exposure on attitudes and behaviors 
is a longstanding research problem~\cite{schmidt2015advertising}.
Despite an extensive body of research, there is no consensus on the exact
relationship; studies have suggested linear~\cite{lee2010exploring, 
schmidt2015advertising}, quadratic~\cite{lee2010exploring, 
chatterjee2003modeling, cox2002beyond, schmidt2015advertising},
cubic~\cite{lee2010exploring}, radical~\cite{deighton1994radical},
logarithmic~\cite{hassan2021logtruth, manchanda2006banner, simon1982adpuls},
and various discontinuous relationships~\cite{betts2019taking, 
yaveroglu2008advertising, chae2019wearout}.
Complicating the issue, these models are often dependent on several covariates
such as brand familiarity, product category, and advertising medium, and
various advertising traits.

A large body of work argues the existence of ``wear-in/wear-out'' effects,
where repeated exposure to an ad first increases, then decreases, 
attitudes towards products and use intent~\cite{pechmann1988advertising,
chae2019wearout, schmidt2015advertising}.
Most model this effect with quadratic terms~\cite{lee2010exploring, 
chatterjee2003modeling, cox2002beyond, schmidt2015advertising} though some
model the phenomenon as an \textit{inverted U}~\cite{lehnert2013advertising}.
Following this body of work and based on preliminary data exploration, we
explore whether the relationship between repeated exposure to \vpnads and
user perceptions follow this inverted U relationship.

Although there has been extensive research on traditional ads (TV ads,
banners, etc.), empirical research on influencer ads is limited.
Further, to the best of our knowledge, the impact of influencer ads for \sp products
remains entirely unexplored.
We select our analysis methods knowing that prior work
in advertising relavent, but do not 
expect these methods to explain our data perfectly.

\section{Methods}
\label{sec:methods}

We hypothesize a relationship between a user's exposure to influencer
VPN ads and their belief in what ads convey. Specifically, we design 
our study to answer these key questions:
\begin{itemize}[itemsep=0pt, left=-4pt, label={}]
    \item \textbf{0.} 
    Does \vpnad exposure correlate with VPN brand familiarity?\footnote{A baseline to validate our method. Ads increase brand familiarity~\cite{schmidt2015advertising}}
    \item \textbf{1.} Does \vpnad exposure correlate with VPN mental models?
    \item \textbf{2.} Does \vpnad exposure correlate with internet threat mental models?
    \item \textbf{3.} What are the most common threats people believe VPNs protect against? Do beliefs mirror prior works' findings?
\end{itemize}

This section describes our study design, independent and dependent
variable definitions, and
analysis methodology.
We pre-registered our analysis plan\footnote{\url{https://aspredicted.org/rk8xe.pdf}}
after our preliminary studies but before the final data collection; our final
analysis did not deviate from our original plan.

\subsection{Study design}
\label{sec:study_design}

Our study comprised three stages: (1)~an initial screener,
(2)~a tutorial stage, and (3)~a final questionnaire stage.
Informed by our extensive piloting and preliminary studies
(\S\ref{sec:piloting}), we designed the staging of
questions with multiple optimization goals in mind: minimize fatigue
and dropout, while maximizing data quality and comfort in the study.
\autoref{tab:data_collected} summarizes our major data sources
and in which stage we collected them.
The screener and tutorial stages were separated by hours to two days,
while the tutorial and final questionnaire stages were separated by
however long it took participants to obtain the relevant data 
(in practice, minutes to 15 days).

\new{Participants were recruited from Prolific. We aimed for a minimum of
\$12/hour (well above U.S. federal minimum wage 
and consistent with Prolific recommendations\footnote{\url{https://researcher-help.prolific.com/hc/en-gb/articles/4407695146002-Prolific-s-payment-principles}})
for time spent on tutorials and surveys.
Because some tutorials were shorter than others, in practice 
some participants received a higher hourly rate (5 mins max, \$1.00 per tutorial). 
We paid for data separately (\$5.25), 
totaling \$9.52--\$10.52 per completed submission. 
Further, we gave bonuses to participants who spent time resolving technical 
issues and paid data compensation to participants who made an effort 
even if their data upload was ultimately unsuccessful.}

Before each stage, participants were given an overview of the 
the current and future stages.
All procedures (detailed below) were approved by 
the University of Maryland Institutional Review Board (IRB).

\begin{table*}[h]
    \centering
    \small
    \begin{tabular}{ p{3.1cm} p{10.9cm} l }
    \toprule
    \textbf{Data} & \textbf{Explanation} & \textbf{Collection stage} \\
     \midrule
        \multicolumn{2}{l}{\underline{\textit{Through questionnaires:}}}\\
    Privacy sensitivity & IUIPC-8 scores. & Screener/tutorial \\
    Hardware configurations & Computer and mobile phone OS & Screener \\
    VPN mental models & Threat, \misleadingshort, \factualshort, all \allmm & Final \\
    Brand familiarity & How familiar participants are with certain brands & Screener/tutorial \\

    \noalign{\medskip}
    \multicolumn{2}{l}{\underline{\textit{Through tutorials:}}}\\
    \yt watch histories & Google Takeout \yt histories, later augmented with video subtitles and details. & Tutorial/final\\
    App download histories & Android or Apple app download histories. Searched for VPN applications. & Tutorial/final\\
    Computer program list & List of programs on Windows or MacOS device. Searched for VPN applications. & Tutorial\\

    \bottomrule
    \addlinespace[1.0ex]
    \end{tabular}

        \caption{Major data sources collected from participants, grouped by collection 
    technique (self-report vs not).} 
    \label{tab:data_collected}
    
\end{table*}

\subsubsection{Screener stage} 

We recruited from gender balanced Prolific users who were 18 years or
older, lived in the US, and used YouTube at least once a month.
The study was advertised vaguely as ``Internet Perceptions Study,'' and
consent was obtained just for the screener intentionally to keep the
real purpose vague (described later in this stage). 
A second formal consent with details of the study was obtained before
the tutorial stage.
We chose this two-tiered consent approach to (1) alleviate selection
bias and (2) to be able to measure the privacy sensitivities of
participants who ultimately could or would not continue.

After the screener consent, participants were asked
questions on their device and YouTube usage.
A small group (n=138) of participants were also asked to complete the
tutorial stage's questionnaire (this group skipped this
questionnaire later).
We use this group to compare the privacy sensitivities of
participants who did and did not complete the study. 
The group was limited since was impractical and statistically 
unnecessary to obtain these from every screener participant.

Finally, we described the full study to participants and asked if
they were willing to share the data that we needed for the study. 
Participants were able to select however many data sources they
were willing to share, but it was clear that those who
did not give all three sources would not continue past the
screener.
Though not required by our ethics committee, this question is
essential to the consent structure of our study.

We admitted any participant to the tutorial stage who met the following
criteria:
(1)~reported to use an Android or iOS primary mobile device, 
(2)~reported to use a Windows or Mac primary computer, 
(3)~watch at least three videos a week on average,\footnote{All
YouTube-related questions were asked for the last year.} 
(4)~``frequently'' or ``always'' be signed in when watching YouTube,
(5)~be ``rarely'' or ``never''~\footnote{All options: ``never,''
``rarely,'' ``sometimes,'' ``frequently,'' ``always.''} not watching
when others were watching YouTube videos through their accounts (this
filters out children who use parents' devices, communal TVs signed into
one account, etc.), 
(6)~have had their YouTube history on for at least the past year, and 
(7)~consented to sharing their data (Android or Apple account app
download histories, installed program list on their primary computer,
and \yt histories).
Participants who did not qualify were paid for their time.

At this stage, participants' hardware configurations and data-sharing statuses 
were recorded automatically on a data collection server we 
instrumented with custom software.
We describe the data collection infrastructure in
\S\ref{sec:tutorial_stage}. The full set of questions 
is in~\autoref{app:screener}.

\subsubsection{Tutorial stage} 
\label{sec:tutorial_stage}
Participants who chose to 
continue with the study were asked for consent 
once again with full study details. Once consent was
obtained, participants first completed a questionnaire consisting 
of IUIPC-8, questions on VPN use (have they used a VPN, what purpose, 
and the brand names), their familiarity with the most popular VPN brands 
(determined by our exploratory data and asked on 
Likert-type options: ``Not at all familiar (1)''--``Very familiar (7)''), 
and an open-ended question asking participants to list the top two 
threats VPNs protect against to the best of their knowledge. Those who 
finished the questionnaire were redirected to the tutorials. 

The tutorial stage consisted of multiple sub-tutorials guiding participants to
provide data or start the exports necessary to provide data later 
in the final stage. The system used hardware configurations 
of participants (collected during the screener) 
and dynamically populated the necessary tutorials. Participants, thus, 
followed multiple flows. Sub-tutorials were combined where 
possible (e.g., Android app
download and YouTube histories). Generally, participants were asked to: 
(1) Start the export of their YouTube histories via 
Google Takeout,\footnote{\url{takeout.google.com}} 
(2) Start the export of their app download 
histories (Google Takeout or privacy.apple.com), 
and (3) export the list of programs on their primary computer.\footnote{Mac users were asked to copy the content of the application 
folder and paste to a text-box in our upload portal. Windows users were 
asked to export relevant windows key registry entries with a bundle of 
PowerShell commands. Again they pasted the exported list to the portal text-box.}

Once tutorials were over, participants were asked to upload any data that 
had already finished exporting, frequently the case for YouTube and Android 
app download histories. Participants whose data had not exported were instructed 
to notify researchers when the data was ready. Communication with participants 
was handled through the Prolific messaging system where no PII is needed.

Participants were finally informed about future procedures. Those 
who did/could not upload all data sources were instructed to notify us 
when data was ready for future tasks. Details about 
the tutorial stage are in~\autoref{app:tutorials}.

\paragraph{Data collection server \& upload portal}
In order to maintain state between stages and ensure 
minimal exposure of participants' data to 
third parties, we developed our data collection server and upload portal. 
Due to the relatively sensitive nature of the data collected, both 
pieces of software featured multiple \sp measures.

The upload portal operated on the client side and 
provided an interface 
to facilitate uploads to the data collection server 
(list of programs on computers, app download histories, 
and YouTube watch histories) and communicated
what data sources we needed from participants 
(see~\autoref{fig:upload_portal} in~\autoref{app:final_stage}). 
The portal only retained files that were 
relevant to the study based on file names 
(history files and program lists). Retained data, along with a 
random 192-bit integrity nonce, was encrypted 
with researchers' public key before they were uploaded 
(we used an audited JavaScript port of 
NaCl~\cite{bernstein2009cryptography}\footnote{\url{https://github.com/dchest/tweetnacl-js}}). 
The corresponding private key was only 
stored on two researchers' laptops. If participants attempted to upload 
a zip with relevant files, 
the portal parsed the zip and discarded any files we were not looking 
for (e.g., Apple exports contain data on physical 
addresses) before encryption and upload.

The data collection server (1)~stored the uploaded encrypted data,
(2)~kept a tally of who signed up for the study (with anonymous
prolific identifiers), and (3)~stored participants' hardware
configurations (Android, iOS, or ``other'' for mobile; Windows, Mac, or
``other'' for computers). All history and program data were received
and stored in an encrypted state. We rate-limited the data collection
server to prevent scans of the prolific ID space or attacks on our
storage capabilities. We logged requests made to our endpoints along
with associated prolific IDs for debugging and forensics purposes.

Researchers downloaded participants' data locally, 
decrypted it (with the private key stored only on two researchers' 
laptops), scraped additional data 
(video subtitles, engagement statistics, upload date, and genres) 
from YouTube.com and its API,\footnote{\url{https://developers.google.com/youtube/v3}}
and analyzed it on their locally. The scraping 
tools are re-implementations of those described 
in~\cite{akgul2022vpn}.

\subsubsection{Final stage}
\label{sec:vpn_final_stage}
Once we received all data sources from participants or received 
confirmation (through Prolific messages) 
that exported data was ready to be provided, and responses/data
passed our quality control tests (discussed below),
participants were admitted to the final stage. 
This stage started with a prompt to upload any remaining 
data sources via the upload portal 
and continued with the final questionnaire. Participants who did 
not provide all required data sources were not allowed to continue. 
The rest were automatically directed to a questionnaire that
contained a series of VPN-ad-related questions 
(Have you heard of VPNs before, where; Have you seen ads for VPNs before, 
where; and an open ended question on what they remember being advertised), 
our main mental models questions (the basis 
for construction of our main dependent variables, 
described in~\autoref{sec:vpn_mental_models}),
and demographics.
We included two attention checks in the mental models 
questionnaires.  Mental model questions were posed 
at the very end to limit bias in the 
earlier open-ended questions, where participants were 
asked to recall their knowledge. 
This arrangement is also likely to increase the quality 
of mental model measurements, as participants' thoughts 
are likely fresh in minds. A full list of questions
is in~\autoref{app:final_stage}.

\subsection{Piloting and preliminary study}
\label{sec:piloting}

In order to develop our mental models questionnaires, 
finalize our study design, and pick a reasonable 
modeling approach to the data we would collect, we conducted a series 
of pilots and preliminary studies. 

\paragraph{Piloting} 
We piloted our study with three usable security 
researchers and three lay-users while developing procedures. 
These pilots helped us gauge the usability of our data collection and 
run end-to-end tests of the data collection infrastructure with various 
participant device configurations.

\paragraph{Mental models questionnaire development} 
We developed two 
questionnaires to measure mental models relevant to our research goals.
After developing a set of statements for each questionnaire based on 
observations from~\cite{akgul2022vpn}, following methodology 
from prior researchers~\cite{votipka2020building}, we collected a series 
of responses from Prolific participants to (1) understand if 
participants were correctly interpreting our statements, (2) eliminate 
ceiling or floor effects, and (3) ensure that our statements were internally 
consistent. We asked open-ended questions after mental models questionnaires
to gauge if participants were correctly interpreting our statements.
After our iterations, we tweaked our Likert options, reworded some statements, 
and added attention checks. Our final scales did not exhibit ceiling or 
flooring effects and had acceptable internal consistency (Cronbach's $\alpha$ 
of .87, .84, .83, and .74)~\cite{krippendorff2004reliability}.

\paragraph{Preliminary study} 
After scale development and piloting, 
we collected a preliminary set of responses from 36 Prolific participants. 
Using this set, we explored different analysis techniques, 
finalized our definition of exposure to \vpnads, 
picked the covariates to consider in our analysis, 
and then pre-registered our exact data collection and analysis procedure.

\paragraph{Data quality control} 
Due to multiple dishonest and careless 
responses during our piloting and preliminary stages, we implemented a series of 
quality control checks throughout our study. Participants had to give 
sensible responses to open-ended questions, pass attention checks, 
give quality data (YouTube histories consistent with survey responses 
(at least a year of data with no obvious gaps in the last year), non-empty 
list of installed programs, and non-empty app download histories). 
Participants who failed any of these checks were not included in the final 
dataset but often received partial compensation based on their time.
Further, any participant we could not find 
influencer VPN ads (in YouTube histories) for but had self-reported to 
have seen 10 or more influencer VPN ads on YouTube in the last 
year were filtered out (n=3). We deemed that these participants had either 
given us missing data or were untruthful with survey responses. 
To incentivize completion, participants were paid for their 
time spent during the different stages of the study but only paid for the 
data they gave upon successful completion of the full study.

\subsection{Ethical considerations}
\label{sec:ethics}
\new{Our research objectives necessitate 
precise measurements of participants' exposure to VPN ads and various covariates 
that could influence mental models. We considered two alternatives when 
collecting this data: purely self-reported data, or observational data}.

\paragraph{\new{Watch histories}} \new{For ad exposure, we preferred the latter. 
Dispite correlation with observational data self-reported 
ad exposure is highly error-prone~\cite{vavreck2007exaggerated, romberg2020validating}. 
This margin of error would inhibit our nuanced analysis. 
Moreover, it is possible that 
ads may impact user mental models even if the users do not actively 
perceive it, let alone recall it years later.}
\new{We collected participants' YouTube watch histories 
in order to measure their exposure to VPN ads. We asked participants to 
share histories from their personal accounts and screened out participants who shared 
their account with others.}

\paragraph{\new{App download histories and computer programs}} 
\new{For a clearer link between mental models and VPN ad exposure, 
we tried to limit major confounds in our experimental setup. 
We suspected that previous use of VPNs might substantially
influence mental models related to VPNs. In our preliminary 
study, self-reported VPN use resulted in less well fitting
regression models than measured VPN use (from App installs and 
computer programs). 
\new{This observation held true in the final data as well, including brand 
familiarity models.}
As with YouTube histories, participants
were instructed to only share data from their personal 
devices---reducing the likelihood of infringing on non-consenting 
parties' privacy. For similar reasons, we did not ask for data from work 
devices. In hindsight, our analysis did not find owning a VPN to be statistically 
significant factor. Future work may chose to omit this variable.}

\paragraph{\new{Protective measures}} Due to the relatively sensitive nature of
the data we collected, we took extra precautions to ensure participants were 
explicitly and implicitly aware of what information they were \new{providing}. We 
obtained formal informed consent for the screener and the full study. During 
the screener, after full procedures were described, participants 
were explicitly asked which data sources they were willing to share and 
whether they would like to continue with the study. We screened out 
participants who reported sharing 
their accounts with others and only asked about personal devices. When 
collecting data, we purposefully designed procedures such that 
participants had to follow tutorials that clearly stated what they 
\new{were} exporting and required them to manually give us the data, 
implying additional consent.
Further, study stages were set up as \new{independent} tasks on Prolific, 
and each task was advertised 
with procedures and expected compensation, enabling reconsideration 
\new{at any stage}. 
One researcher actively monitored data collection and 
helped debug any issues that came up during the study. We compensated 
participants for extra time spent if we deemed their efforts to be made 
in good faith. To reduce exposure of participants' data to third 
parties (including system administrators), we developed our own data collection 
procedures and only kept fully decrypted watch/download history and 
computer program data on two researchers' 
laptops (see~\autoref{sec:tutorial_stage}). 
Procedures were created iteratively, and 
all were approved by the 
University of Maryland IRB.

\begin{table*}[h]
    \centering
    \small
    \renewcommand{\arraystretch}{1.1}
    \begin{tabular}{ p{4.1cm}ll}
    \toprule
    \textbf{Variable} & \textbf{Explanation} & \textbf{Details} \\
     \midrule
        \multicolumn{2}{l}{\underline{\textit{Dependent variables (DVs):}}}\\
    Factual VPN mental models & (Dis)agreement with factual statements about VPNs featured in \vpnads. & \autoref{sec:vpn_mental_models} \\
    Misleading VPN mental models & (Dis)agreement with misleading statements about VPNs featured in \vpnads. & \autoref{sec:vpn_mental_models} \\
    All VPN mental models & (Dis)agreement with statements featured in \vpnads. & \autoref{sec:vpn_mental_models} \\
    Threat mental models & (Dis)agreement with threat statements featured in \vpnads. No mention of VPNs. & \autoref{sec:vpn_mental_models} \\

    VPN brand familiarity ($\times5$) & Brand familiarity with ExpressVPN, NordVPN, ShurfShark, PIA, and Atlas VPN. & \autoref{sec:additional_variables} \\

    \noalign{\medskip}
    \multicolumn{2}{l}{\underline{\textit{Independent variables (IVs):}}}\\
    Exposure to VPN ads & A measure of how much exposure to influencer VPN ads.& \autoref{sec:measuring_exposure}  \\ 
    VPN ownership & VPN software found in the list of programs or apps (didn't own). & \autoref{sec:additional_variables} \\
    Technical expertise & How often participants are asked for tech advice. Bucketed into two (less often). & \autoref{sec:additional_variables}\\
    Privacy sensitivity & Participants' IUIPC-8 scores. & \cite{malhotra2004internet, gross2021validity} \\
    VPN ad interval & Average time between VPN ads. & \autoref{sec:additional_variables} \\
    \bottomrule
    \addlinespace[1.0ex]
    \end{tabular}
    \caption{Independent and dependent variables (IVs, DVs) used in analysis. Categorical variable baselines in parentheses.} 
    \label{tab:vpn_variables}
    
\label{tab:ivs}
\end{table*}

\subsection{Measuring and defining variables.}

Here we define the main independent and dependent variables used in our analysis, 
summarized in~\autoref{tab:vpn_variables}.

\subsubsection{Mental models questionnaires}
\label{sec:vpn_mental_models}

\begin{table*}[h]
    \centering
    \small
    \renewcommand{\arraystretch}{1.1}
    \begin{tabular}{ rll}
    \toprule
    \textbf{Scale group} & \textbf{Staement} \\
     \midrule
    \multicolumn{2}{l}{\underline{\textit{VPN mental models:}}}\\
    Misleading & I don’t have to worry about anything on the internet if I use a VPN. \\
    Misleading & Companies can’t collect my data on the internet if I use a VPN. \\
    Misleading & My credit card information is protected online if I use a VPN. \\
    Misleading & My passwords are protected online if I use a VPN. \\
    Misleading & I am protected from seeing ads on the internet if I use a VPN. \\
    Factual & My Internet Service Provider (e.g., Verizon, AT\&T) can’t find out which websites I go to if I use a VPN. \\
    Factual & Hackers on the same wireless network as me can’t see which websites I go to if I use a VPN. \\
    Factual & I can watch other countries' streaming libraries (e.g., Netflix, Hulu) if I use a VPN. \\
    Factual & I can overcome internet censorship if I use a VPN. \\
    Factual & It seems like I’m browsing the internet from somewhere else, if I use a VPN. \\
    Factual & A VPN encrypts my web traffic before it leaves my device. \\
    \multicolumn{2}{l}{\underline{\textit{Threat mental models:}}}\\
    Threats & My Internet Service Provider (e.g., Xfinity, AT\&T) is tracking all of my internet activity.  \\ 
    Threats & My Internet Service Provider (e.g., Xfinity, AT\&T) is selling all of my internet activity. \\
    Threats & The U.S. Government is tracking everyone (including me) online. \\
    Threats & Internet companies are collecting all of my internet activity. \\
    Threats & Hackers can easily steal my credit card information from me online. \\
    Threats & Hackers can easily steal my passwords from me online. \\

    \bottomrule
    \addlinespace[1.0ex]
    \end{tabular}
    \caption{Likerts used for measuring mental models of threats and VPN capabilities. Item order was randomized for participants.} 
    
\label{tab:statements}
\end{table*}

We measure participants' mental models by administering two
Likert-type questionnaires: one for understanding 
of what VPNs are capable of, and a 
broader one for threats on the internet. Unlike measurements of exposure to 
\vpnads, using questionnaires to consistently measure mental models is still 
one of the only feasible 
solutions~\cite{story2021awareness, akgul2021evaluating}. Both scales were based 
on the most prominent observations from our previous work~\cite{akgul2022vpn} but
also ensured adequate coverage of the space of observed statements 
(e.g., we did not keep both ``Governments are tracking you'', and 
``Governments are tracking everyone,'' the two most popular themes in threats). 
The full set of statements along with what it measures are in~\autoref{tab:statements}.

\paragraph{VPN mental models} Following observations from prior work, 
the first questionnaire attempted to measure a combination 
of prevalent statements about VPNs in \vpnads, as well as more specifically 
factual and misleading ones. These statements were all about VPNs directly, 
were presented together due to contextual relevance, 
and were asked using prevalence-based Likert-type options of 
``True for all VPNs,'' ``True for almost all VPNs,'' 
``True for some VPNs,'' ``True for almost no VPN,'' and
``True for no VPN.'' We chose this framing as opposed to
``Agree/Disagree'' options because a large minority of participants interpreted 
agreement as prevalence during 
piloting and preliminary studies, while nearly all correctly interpreted the 
prevalence options.
Scales were created out of Likert responses 
simply by adding up numeric 
values equivalent to the position of the 
selected Likert-type option (e.g., ``True for all VPNs,'' $\rightarrow$ 5, ``True for no VPN'' $\rightarrow$ 1). 
The resulting scales for all mental models, misleading mental models, 
and factual mental models
had Cronbach's alphas of .87, .84, and .74 
respectively in a set of 88 preliminary Prolific responses. We deemed 
these to be acceptable, with results on factual mental models considered 
relatively tentative.

\paragraph{Threat mental models} The second questionnaire measured 
agreements with broader \note{(hyperbolic)} threat statements 
that were featured in \vpnads but were not specifically about VPNs 
\new{~\cite{akgul2022vpn}}. 
We measured these threat statements separately from VPN-related ones 
since mental models of threats might have 
implications for security \& privacy at a broader level. 
These statements are not necessarily tied to any specific technologies; thus, 
we asked participants'
agreement using a seven point 
Likert from ``Strongly agree,'' to ``Strongly disagree.'' We created 
a \note{six-item} scale (see~\autoref{tab:statements}) and obtained a Cronbach's alpha of .83, 
once again, acceptable~\cite{krippendorff2004reliability}.

\note{Although the threats we use in this scale contain some truth, we consider them hyperbolic: 
(1) ISPs track internet activities and sell derived data products; however, they cannot track \emph{all} 
activity as most content is encrypted~\cite{GoogleTransparency2024}. 
(2) Though it is hard to know exactly what the U.S. government does, 
the fourth amendment (and related statutes~\cite{fourth,fisa}) does constrain surveillance to varying degrees~\cite{eff_privacy2024}. (3) Due to end-to-end encrypted 
communications (most popular messaging apps) and data privacy measures, internet companies 
cannot collect \emph{all} data. (4) To steal passwords and credit 
cards, hackers need to implement nuanced attacks, often requiring physical 
proximity (e.g.,~\cite{vanhoef2017key}), or sophisticated attacks 
on identity/transaction management systems (e.g.,~\cite{Rockefeller2014}). 
This limits the effectiveness of such attacks, making 
them rare for the average user~\cite{breen2022large}.
The threats mentioned are real problems, and we don't intend to 
minimize them here, but these scale items --- and the ad contents they are 
drawn from --- use language (e.g., all, everyone, easily) 
in ways that we consider hyberbolic.}

\subsubsection{Measuring exposure to ads}
\label{sec:measuring_exposure}

\begin{figure}
    \centering
    \includegraphics[width=.8\linewidth]{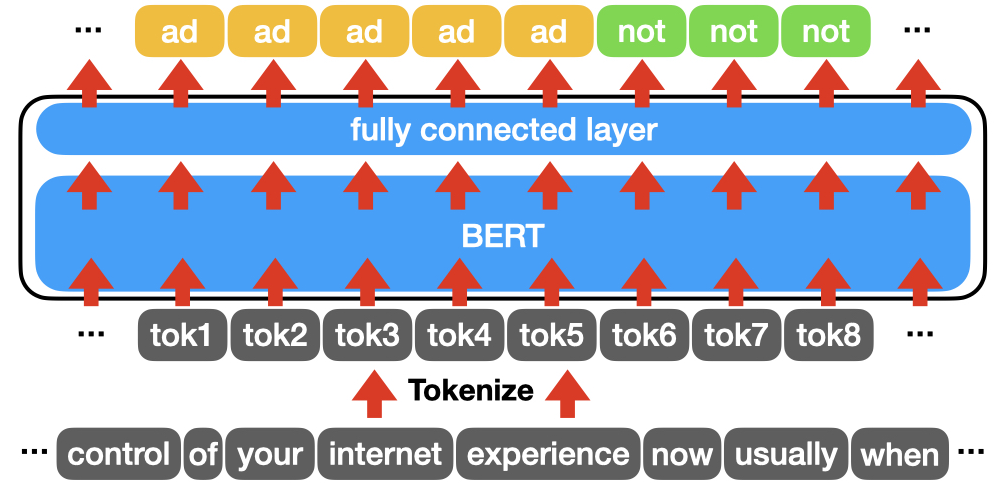}
	\caption{Ad classifier overview. In this example, 
	``now usually when'' is not part of an ad, the rest are.}
    \label{fig:bert}
\end{figure}

The second component we must be able to measure is our participants' 
exposure to influencer VPN ads. We base our exposure calculation on 
the total number of words that make up influencer 
VPN ads in the \yt histories of our participants.
We use the amount of words, as opposed to individual ad 
segment counts or aggregate duration of all influencer VPN ads, because 
we believe word count is a more accurate measure of information conveyed; 
some ads might be longer than others and some \yters 
might produce denser, faster-paced content compared to their peers.

To count VPN ad words, we augment BERT~\cite{devlin2018bert}
with a fully connected output layer to produce an output 
vector of predictions given an input sequence of 
tokenized text (in our case, \yt video transcripts). 
The objective is to output predictions that identify whether the 
corresponding input token (i.e., words or word parts) 
is part of a VPN ad or not. To parse videos that contain 
more than 512 tokens (a BERT limitation), we use a sliding window 
with a stride of 256, classifying tokens as part of an 
ad if either of the two predictions classify it as such. Once 
each word in a video is labeled, the problem reduces 
to a count of the ad-classified tokens. An overview of our approach is 
given in~\autoref{fig:bert}.

To train our model, we use two datasets of video subtitles, 
each labeled per subtitle token as part of a VPN ad or not. 
The first dataset contains a random sample of 
238 VPN ad and 238 non-VPN-ad videos (along with labels for where the ads are), 
from our prior work~\cite{akgul2022vpn}.
The second dataset was obtained by manually labeling 351 VPN ads from a 
crowd-sourced dataset of influencer \yt 
ads\footnote{\url{https://sponsor.ajay.app/database}}. Two coders labeled 
videos following the convention we previously set~\cite{akgul2022vpn}, identifying 
relevant videos with keyword searches, reaching 
agreement\footnote{We reached a Krippendorff's $\alpha$ of 
.98 over 300 ad segments, 
($\alpha=.97$ over 165 videos), 
greater than the acceptable 
.85
from our previous work.}, 
and coding additional videos individually.
We use the same dataset to include 351 non-VPN influencer ads in 
our dataset to ensure our classifier does not confuse such ads with \vpnads, 
a common pitfall in early experimentation. 
The overall dataset included labeled segments for both \textbf{short} 
(introductory or transient; e.g., ``and a big 
thank you to nordvpn for sponsoring 
this video. If you liked [this] video 
make sure to smash that like button'') 
and \textbf{main} (about a minute long of 
describing threats online and benefits of VPNs) VPN segments 
as observed 
in our prior work~\cite{akgul2022vpn}.

\paragraph{Evaluation} We calculate several metrics for model assessment, 
including traditional metrics and a qualitative 
approach.

We first evaluate the model with five-fold cross-validation. 
On a per-word label basis, we achieve precision, recall, and f1 
scores of .89, .88, and .88 respectively.   
Calculating the same metrics for each video 
in the validation set, we get an average 
precision, recall, and f1 of 0.92 ($\sigma=.20$), 0.96 ($\sigma=.11$), 
and .92 ($\sigma=.21$) respectively\footnote{As in prior work, we modify 
metrics to mitigate zero division 
errors~\cite{usbeck2015gerbil}.}.
If we use any words' positive labeling within a video 
to equate to the video containing an ad, we achieve ad detection 
precision, recall, and f1 scores of 0.99, 0.94, and 0.97.

Next, to get a sense of how well this model would work on 
our participants' history dataset and to understand where it 
might not work as well, we take a semi-qualitative approach.
We manually evaluate the model's predictions on a random sample 
of up to five videos per exploratory stage participant 
with at least one VPN-related phrase 
(e.g., ``VPN,'' ``SurfShark,'' ``virtual private''), 
adding up to 138 unique videos. 
We call this set the \textit{measurement validation set}.
Two researchers independently searched through the videos and tried to 
identify \vpnads by (1)~looking at 
common places where \vpnads appear, 
(2)~keyword searches, and (3)~the model inference results. 
They assigned one of two labels per ad segment 
to establish a ground truth set, 2 and 1,
indicating the segment was in its entirety a VPN ad, 
or a VPN was part of the advertised product (henceforth, partial ads). 
We chose this qualitative way of labeling videos since the 
boundaries of \vpnads are often 
fuzzy and we wanted to differentiate 
the model's success on short segments compared to main segments. 
This style offered consistency with labels 
between researchers. The researchers additionally 
noted if the segments they labeled were main ad 
segments (usually a minute long, with the 
bulk of the content~\cite{akgul2022vpn}) 
or short segments that were meant to remind viewers of the sponsor, 
or briefly mention the VPN product.
They then repeated this process to judge all 
segments detected by the model. On participants' 
data, our model achieves precision, recall, and f1 
score of .87, .90, and .88 respectively.\footnote{Due to our labeling style, 
regular VPN ad segments contribute twice as much to the scores (positively and negatively) 
as partial ads segments.} 

Classifiers intended to detect \sp content are 
rare, making it hard to establish a fair baseline.
However, we note that these numbers are arguably 
better than 
our previously reported
agreement when 
we
manually labeled videos for VPN ads (Krippendorff 
$\alpha=.85$,~\cite{akgul2022vpn}), and on par with 
classifiers attempting to label privacy content with 
much larger language models (precision$=.91$, 
recall$=.84$~\cite{akgul2024decade}). 

\paragraph{Misclassifications}
We deem our models to be accurate enough 
for the type of analyses we plan to 
conduct. Regardless, misclassifications 
might introduce certain biases to our results. 
We notice three ad segment types that constitute 
nearly all misclassifications in the manual validation set: 
(1)~short segments that usually do not contain much \sp-relevant 
content (16/32); (2)~not being able to detect partial ad segments (6/32); 
and (3)~segments that praise
VPNs but are not explicitly sponsored (7/32). Notably, our metrics are 
significantly higher for the main ad segments, where 
most of the \sp relevant content is disseminated 
(recall$=0.98$, precision$=.90$, 
f1$=.94$). Thus, most misclassifications (short segments, or partial ads) 
likely only impact our analysis 
of brand familiarity.

\paragraph{Defining exposure} We define exposure to influencer VPN ads 
for participant $i$ as:
\small
\[\mathrm{ad~exposure}_i=
	\ln\left(\frac{\mathrm{VPN~ad~word~count}_i}
			{ \frac{1}{N}\left(\sum^{N}_{k=1} \mathrm{VPN~ad~word~count}_k \right)} +1 \right)\]

\normalsize

\noindent
where $N$ denotes the number of videos (among all participants) 
that contain at least one ad-word. 
We use a log scale to account for the diminishing returns of repeated ads---the 
first watch will not necessarily have the same effect as the 50th. 
The utility of a logarithmic scale was observed in early
ad response work~\cite{rao1972logsads} and has been
adopted in various formulations of ad exposure 
since (e.g.~\cite{manchanda2006banner, danaher2013comparing, chae2019wearout}).
Further, we had empirically observed (on preliminary data) 
that a log scale results in more accurate modeling 
on our exploratory data. The ``$+1$'' is to avoid undefined values
for participants who do not have VPN ads in their histories.

\subsubsection{Additional variables}
\label{sec:additional_variables}
Our preliminary analysis indicated 
that more than just exposure to ads might 
be correlated with mental models. 
These variables had either been found to 
potentially be correlated with mental models (p$<$.10) in our 
preliminary study or prior work.
Specifically, we included the following covariates.
We considered including demographic variables; 
however, aside from technical background, 
our preliminary study found no relationship between demographics and mental models.

\paragraph{Tech savviness} We asked participants how 
often they gave tech advice, as a proxy for 
tech savviness~\cite{akgul2021evaluating,redmiles2016learn}. 

\paragraph{Privacy sensitivity} Privacy sensitivity is likely 
to be associated with participants' understandings of VPNs and 
threats. We measure privacy sensitivity with 
IUIPC-8~\cite{malhotra2004internet,gross2021validity}, a widely accepted 
privacy scale.

\paragraph{VPN ownership} Our preliminary analysis 
showed VPN ownership history to correlate with mental models.
\new{We explored two measures of ownership: self-report and measured. 
Self-report data resulted in weaker-fitting models in the preliminary 
dataset; thus we used the measured 
approach (this is true for the final dataset, 
including brand familiarity models).}

\new{For completeness, we also explored 
including self-reported use of VPNs in work contexts in 
additional to personal use. This degraded the fit of the 
regression models in both the preliminary and final datasets, 
including brand familiarity models.}
\new{Ultimately, we chose to focus on consumer VPNs 
(instead of corporate solutions) because 
(1) consumer VPNs were the focus of VPN ads~\cite{akgul2022vpn}, 
(2) corporate VPNs may conduct surveillance or other privacy-invasive features (e.g.,~\cite{ciscoSurveillance}), 
(3) and corporate VPNs might be enabled by default 
(e.g., on \emph{managed} devices~\cite{apple2024managing, appleVPN, microsoftVPN}) 
without users realizing.} 

\new{Since the most popular VPNs that advertised in our dataset are 
used via native apps},\footnote{\new{Exact numbers 
are inaccessible but, these apps seemingly receive an order of magnitude more 
downloads than extension installs. E.g.,} \href{https://chromewebstore.google.com/detail/expressvpn-vpn-proxy-for/fgddmllnllkalaagkghckoinaemmogpe}{1M ExpressVPN extension
downloads} vs \href{https://play.google.com/store/apps/details?id=com.expressvpn.vpn}{50M+ ExpressVPN app installs}}
we search the app download histories of participants' primary 
Android or Apple devices along with the 
list of programs installed on their 
primary computers, and consider a 
participant to have owned a VPN if 
we find a program that has ``VPN,''
or any of 37 popular VPN brands (found in prior work, in online 
ranking lists) in the title.

\paragraph{Average VPN ad interval} Prior work on 
traditional ads had 
found that the effectiveness of 
ads depends on how frequently customers 
are exposed to ads 
(e.g.,~\cite{schmidt2015advertising, malaviya1997persuasive}). We measure the 
average days between viewing \vpnads for each participant using their 
\yt watch histories.

\paragraph{\new{Brand familiarity}} \new{Prior work has consistently found a 
link between brand exposure and brand familiarity~\cite{schmidt2015advertising}.    
We explore the correlation between VPN ad exposure 
to specific brands and familiarity with that brand as a baseline check on the quality of our data. 
Using Likert-type scales (``not at all familiar'' to ``very familiar''), 
we measure familiarity with the five most popular brands from the preliminary study.}

\subsection{Analysis}
\label{sec:analysis}
Our goal is to investigate the 
relationship between ad exposure and mental models. 
As such, our main analysis consists of regressions on all four mental models 
variables (\factualshort, \misleadingshort, \allshort, and \threatshort mental
models). 

We do not try to model the exact shape of the relationship 
curve, as there is no clear consensus in the marketing literature 
on what the ad response function is (see~\autoref{sec:modeling_ads}, modeling ads). 
There seems to be consensus around 
wear-in/wear-out effects, and our 
preliminary analysis suggests the same.
Quadratic terms have often been used to model this 
effect (\cite{lee2010exploring, chatterjee2003modeling, cox2002beyond, 
schmidt2015advertising}), but
heavily more recently~\cite{simonsohn2018two}. 
Thus, we simply test if the relationship 
between exposure to ads and mental models starts positive and 
turns negative after the maximum. 

To achieve this, we specifically 
conduct \textit{two lines} analysis proposed by 
Simonsohn~\cite{simonsohn2018two} once per mental model variable. After 
finding the maxima, we fit two additional linear regression 
models per mental model variable to account for covariates 
(privacy sensitivity, tech savviness, VPN ownership, 
and average VPN ad interval): one before the maximum and one 
after. We do not include covariates in the two 
lines analysis since covariates might 
have different effects before/after 
the maxima, this in unaccounted for by two lines analysis.

We choose to include the following covariates in
the set of independent variables: exposure to VPN ads, 
IUIPC-8, tech advice frequency (bucketed into two), ownership of VPNs, 
and mean time between exposure to ads.
To avoid overfitting, we fit all possible models with combinations of 
these IVs that contain exposure to ads (our main variable of 
interest). We then select the adjusted-$R^2$ maximizing model 
for each DV. This analysis was repeated for each of the four 
mental models.

As a \x{sanity} check on data quality, 
we largely use the same process as 
before but use brand familiarity as 
the dependent variable and exposure 
to the specific brand ad (same definition 
from before, but split by 
brands) as the main covariate. Unlike mental models, 
we do not run two lines analysis, we simply select 
(same selection strategy as before) one ordinal logistic regression per 
brand. We chose this approach 
since brand familiarity doesn't decrease 
with increased ad exposure~\cite{schmidt2015advertising}. 
Following prior work, the IVs were augmented 
with exposure to all other VPN brand ads~\cite{libai2009role}.

We performed qualitative analysis on three short, open-ended questions
in our survey: (1) what threats do VPNs protect against, (2) where did participants 
learn about VPNs, and (3) justifications for a response in the VPN
mental models scale (selected at random for each participant).
For each question, we established adequate inter-rater reliability metrics.
All agreements were calculated over 10\% of their respective datasets.
For the first question (threats), two researchers used our
existing codebook (obtained from~\cite{akgul2022vpn}) 
of VPN and threat statements found in VPN advertisements. Following 
our prior approach~\cite{akgul2022vpn}, we coded the
assets under threat, adversaries, and attacks per response
to establish threat models.
After obtaining a Krippendorff's $\alpha$ of .88 for combined codes 
(Strictly defined~\cite{akgul2022vpn}. Individual subcode 
agreements are .94 for adversaries, .90 for attacks, 
and .97 for assets; all much higher than those 
previously reported~\cite{akgul2022vpn}),
researchers split the remaining responses evenly to code.
Two researchers iteratively developed codebooks 
for the second (information source)
and third (justifications) questions from scratch; after 
agreement was reached (Kupper-Hafner
concordance of .92~\cite{malkin2023optimistic,harbach2016locking} and
Krippendorff's $\alpha$ of .76~\cite{krippendorff2004reliability} 
respectively), one researcher coded all remaining responses.

To account for the sampling 
bias potentially introduced by the demanding 
nature of our study, we collect IUIPC-8 scores from (n=98) 
participants who ultimately did not finish the study. We 
compare the IUIPC-8 of this set and the final 
set of participants with a Mann-Whitney U test.

\subsection{Limitations}
\label{sec:limitations}

We recruit our participants from a crowdsourced sample, limitations of 
which are well known: participants tend to be younger, more educated, 
and privacy-conscious than the general US population. 
Our measurements indicate that our participants might have been slightly
less privacy-sensitive than general users on the platform, though our
participants included a wide range of ages, educational backgrounds, and
had a similar race distribution as the US public.

\new{Our work focuses on influencer VPN ads, created by 
\yters and embedded directly in the video content (as opposed to interstitial \yt ads). 
Participants could have received VPN ads through other means (e.g., TV ads, 
podcasts), or when they weren't logged into their personal accounts. This would 
confound our analysis.}
We screened out participants who often watched \yt while not logged
in, or who had others (e.g. partners)
regularly watch \yt on their accounts in their absence.
Our results show that \yt is the primary platform \vpnads appear
by a large margin \new{(see~\autoref{sec:demo}), indicating 
minimal outside platform influence}.

Our questionnaires might not have distilled mental models
accurately.
We used the most prevalent statements from 
our
representative sample of \vpnads~\cite{akgul2022vpn}, and extensively 
tested our questionnaires (multiple pilots and a preliminary study).

\new{Our detection of VPN use is likely not perfect. Though the most 
popular brands in our dataset all had native applications, some also 
had extensions which our measurement wouldn't detect. The most popular 
VPNs in our dataset are primarily used through native apps, limiting this 
issue (see~\autoref{sec:additional_variables}).}

It is possible that some \vpnads were too nuanced for our classifier
to detect.
Based on our assessments (\autoref{sec:measuring_exposure}), we believe
our model is sufficient for our analysis.

We did not and could not measure exposure to \vpnads outside of \yt.
Similarly, we limit causal arguments as we cannot measure every
variable that might explain VPN and \sp mental models.
However, our purpose-built questionnaires captured the most common claims in
\vpnads, and controlled for major confounding variables.

\new{Our modeling of exposure is likely incomplete. On the preliminary dataset 
we experimented with several measures from prior work, 
ours produced the best models. Our goal isn't 
necessarily to find the best model but a useful one.}

US participants were recruited for this study. 
Our results might not generalize well to populations with different
cultures and societal norms.
This is an important limitation, as a large portion 
of the VPN market is outside of the US~\cite{vpnadoption}.

\section{Results}
\label{sec:results}

We present our results in this section. First, we characterize 
our participants. Then, we present our main analyses.

\begin{table}[t]
    \setlength\tabcolsep{5pt} 
    \renewcommand{\arraystretch}{1} 
    \footnotesize
    \resizebox{1\linewidth}{!}{

    \begin{tabular}{p{2cm} p{3.5cm} p{1cm} p{1cm}}
    \toprule
    \midrule
    \textbf{Gender}         & Female                  &   98 \\
                            & Male                    &   105 \\
                            & Self described          &   14 \\
    \midrule
    \textbf{Age}            & 18-25                 &    53 \\
                            & 26-35                 &    82 \\
                            & 36-45                 &    47 \\
                            & 46-60                 &    25 \\
                            & 61+                   &    10 \\

    \midrule
    \textbf{Ethnicity}      & White                      & 141 \\
                            & Black or African Am.       & 17  \\
                            & Asian or Asian Am.         & 23  \\
                            & Hispanic or Latino         & 12  \\
                            & Other or mixed race        & 24  \\
    \midrule
    \textbf{Education}     
                    & Completed H.S. or below         & 28 \\
                    & Some college, no degree		  & 38 \\
                    & Trade or vocational             & 5 \\
                    & Associate's degree              & 26 \\
                    & Bachelor's degree               & 97 \\
                    & Master's or higher degree       & 23 \\
                
    \midrule
    \textbf{YouTube videos}  & $<$10K            & 50 \\
    \textbf{in history}      & 10K-30K     & 61 \\
                          
                          & 30K-50K     & 92 \\
                          
                          & 50K+     & 14 \\
    \midrule
    \textbf{Give technology} & Always, often              & 73 \\
    \textbf{advice}     & Sometimes, rarely, never     & 144 \\
    \midrule
    \textbf{VPN} & Yes                       &  77  \\
    \textbf{ownership}                    & No                       &  140 \\

    \midrule
    \bottomrule
    \end{tabular}
    }

    \caption{\small Participant demographics.}
    \label{tab:vpn_demographics}
    \end{table}

\subsection{Demographics/participants}
\label{sec:demo}
In total, we screened 2755 participants, 
$\sim$830 of which met our eligibility criteria
and were invited. Ultimately,
217 participants completed our full study in August/September 2023. 

While we collected data across a diverse pool of ages and education levels,
our participants skewed younger and more educated compared to the US
population (see~\autoref{tab:vpn_demographics}).

The privacy sensitivities (IUIPC-8) 
of participants who finished the study were significantly 
lower than those who wished to not continue (p=$0.03$) with a 
location shift estimate of 2.00 (IUIPC-8 range: [7, 56]). 
In contrast, no significant difference was found
(p=$0.07$, location shift estimate: 1.00) 
between participants who finished the study and those who 
were screened out (wished to not continue, 
incompatible hardware, low \yt usage). 
This plausible selection bias might mitigate 
the much higher privacy sensitivities of 
Prolific participants compared to the general 
population~\cite{tang2022replication},~\cite{abrokwa2021comparing}. 

\paragraph{Watching \yt and \vpnads}
The mean length of our participants' YouTube histories  
was 6.42 years (min=1.01, max=13.07, $\sigma$=3.6).
On average, participants watched 
27.00K (min=0.43K, max=90.31K, $\sigma$=16.98K) \yt videos 
with 82.26 (min=0, max=1050, $\sigma$=124.42) containing VPN ads. 

Self-reported responses align with these measurements of 
\yt use and \vpnad exposure. When asked to 
recall (open-ended) where they heard about VPNs, 52.5\% of participants 
volunteered ads, 37.7\% mentioned \yt, and 29.5\% mentioned ads/sponsorships on \yt.
Among participants who said they had seen 
VPN ads embedded in videos (165 out of 217), 160 (97.0\%) 
selected YouTube as the medium, implying that YouTube 
is the primary distributor such ads. 
In contrast, 12.7\% selected Twitter, 
10.9\% TikTok, 10.9\% 
Instagram, 8.4\% Facebook, 
and 12.7\% selected TV.

\subsection{Brand familiarity and VPN ad exposure}
\label{sec:brand_familiarity}

\begin{table*}
    \footnotesize
    \centering
    \begin{tabular}[]{r p{.029\linewidth}p{.075\linewidth} r p{.029\linewidth}p{.075\linewidth} r p{.029\linewidth}p{.075\linewidth} r p{.029\linewidth}p{.075\linewidth} r p{.029\linewidth}p{.075\linewidth}}
    \toprule
    \midrule
      & \multicolumn{2}{c}{ExpressVPN}  & & \multicolumn{2}{c}{NordVPN} & & \multicolumn{2}{c}{SurfShark}  & & \multicolumn{2}{c}{PIA}  & & \multicolumn{2}{c}{AtlasVPN} \\
      & \multicolumn{1}{c}{OR} & \multicolumn{1}{c}{95\% CI}  & & \multicolumn{1}{c}{OR} & \multicolumn{1}{c}{95\% CI} & & \multicolumn{1}{c}{OR} & \multicolumn{1}{c}{95\% CI} & &\multicolumn{1}{c}{OR} & \multicolumn{1}{c}{95\% CI} & & \multicolumn{1}{c}{OR} & \multicolumn{1}{c}{95\% CI} \\
      \cmidrule{2-3} \cmidrule{5-6} \cmidrule{8-9} \cmidrule{11-12} \cmidrule{14-15}
    Brand exp.         & 1.41* & [1.01, 1.96] & & 1.51*** & [1.27, 1.79] & & 2.30*** & [1.68, 3.19] & & 1.98*** & [1.38, 2.86] & & 1.7** & [1.23, 2.35]\\
    IUIPC-8 score      & 1.03  & [0.99, 1.08] & & 1.05*   & [1.01, 1.10] & &         &              & & \\
    Low tech adv.      &       &              & & 0.45**  & [0.27, 0.76] & & 0.98    & [0.54, 1.80] & & 0.91    & [0.47, 1.81] & & \\
    Had VPN            &       &              & & 1.37    & [0.82, 2.27] & &         &              & &         &              & & 0.94  & [0.48, 1.80]\\
    Avg. ad interval   & 1.00  & [1.00, 1.00] & &         &  & & & & & \\
    Non-brand exp.     & 1.19  & [0.82, 1.75] & &         &              & & 1.07    & [0.88, 1.34] & & 0.96 & [0.85, 1.11] & \\
    \midrule

    Ad count      & \multicolumn{2}{c}{7631} & & \multicolumn{2}{c}{3784}   & & \multicolumn{2}{c}{2558}  & & \multicolumn{2}{c}{633}  & & \multicolumn{2}{c}{442} \\
    AIC           & \multicolumn{2}{c}{702.8} & & \multicolumn{2}{c}{795.1} & & \multicolumn{2}{c}{538.6} & & \multicolumn{2}{c}{437.3}  & & \multicolumn{2}{c}{397.2} \\
    R2 Nagelkerke & \multicolumn{2}{c}{0.42}  & & \multicolumn{2}{c}{0.21}  & & \multicolumn{2}{c}{0.31}  & & \multicolumn{2}{c}{0.08}  & & \multicolumn{2}{c}{0.06} \\
    \midrule
    \bottomrule
    \end{tabular}
    \caption{Ordinal logistic regression models predicting familiarity with VPN brands. OR: odds ratios, exp.: \vpnad exposure. 
    } %
    \label{tab:vpn_brand_exposure}

\end{table*}
\begin{table*}
\footnotesize
\centering
\begin{tabular}[t]{r ll p{0.0cm} ll c ll c ll}
\toprule
\midrule
  & \multicolumn{2}{c}{Threats} & & \multicolumn{2}{c}{Factual} & & \multicolumn{2}{c}{Misleading} & & \multicolumn{2}{c}{All VPN statements} \\

  & \multicolumn{1}{c}{Estimate} & \multicolumn{1}{c}{95\% CI}  & & \multicolumn{1}{c}{Estimate} & \multicolumn{1}{c}{95\% CI} & & \multicolumn{1}{c}{Estimate} & \multicolumn{1}{c}{95\% CI} & & \multicolumn{1}{c}{Estimate} & \multicolumn{1}{c}{95\% CI} \\

  \cmidrule{2-3} \cmidrule{5-6} \cmidrule{8-9} \cmidrule{11-12}

VPN ad exposure      & ~0.84*** & [~0.35, 1.33] & & ~0.06  & [-0.24, 0.37]  & & -0.41+ & [-0.87, 0.05] & & -0.22 &  [-0.83, 0.40]\\
IUIPC-8 scores       & ~0.12+   & [-0.01, 0.25] & & ~0.07+ & [-0.01, 0.15]  & &        &               & & ~0.07 &  [-0.09, 0.24]\\
Low tech adv.        & -1.38    & [-3.15, 0.38] & & -0.87  & [-1.97, 0.23]  & &        & \\
Had VPN              & -1.01    & [-2.75, 0.73] & & ~0.99+ & [-0.10, 2.08]   & & -0.80   & [-1.99, 0.40] & & \\
Avg. VPN ad exposure &         &                & &      &                 & & ~0.00   & [-0.01, 0.00] & & \\
\midrule
AIC        & \multicolumn{2}{c}{1402.5} & & \multicolumn{2}{c}{1200.2} & & \multicolumn{2}{c}{1120.6} & & \multicolumn{2}{c}{1515.2}\\
$R^2$      & \multicolumn{2}{c}{0.089}  & & \multicolumn{2}{c}{0.049}  & & \multicolumn{2}{c}{0.026}  & & \multicolumn{2}{c}{0.005}\\
\midrule
\bottomrule
\end{tabular}
\caption{Linear regression models for mental models. +: p $<0.1$, *: p$<0.05$, **: p$<0.01$, ***: p$<0.001$} 
\label{tab:mental_models}
\end{table*}

To start to understand our 
data, we explore the relationship between 
exposure to \vpnads and familiarity with VPN brands via 
ordinal logistic regressions. As described 
in \autoref{sec:analysis} we select the best 
fitting brand familiarity model for each of the 
top five VPNs that appeared in the preliminary data. 
We report confidence intervals for each coefficient and, 
due to its similarity with OLS $R^2$, we 
report Nagelkerke's pseudo-$R^2$~\cite{nagelkerke1991note, walker2016nine}.

\begin{figure}[t]
    \centering
    \includegraphics[width=.90\linewidth]{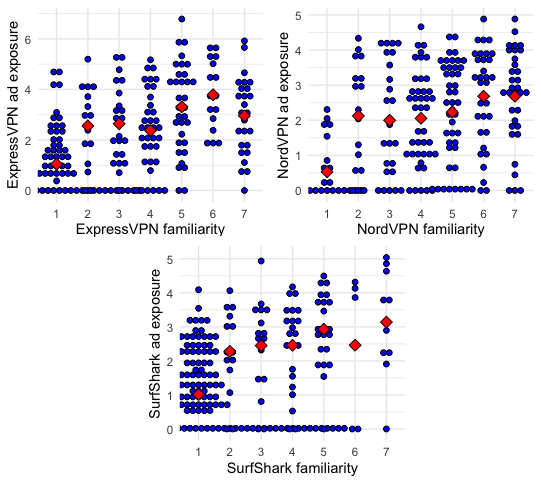}
    \caption{Brand familiarity and VPN ad exposure for each brand. Red diamonds 
    denote the means. Familiarity ranges from ``not at all familiar'' (1) to ``very familiar'' (7).}
    \label{fig:brand_exposure}
\end{figure}

The results show that every order of magnitude increase 
in exposure (defined on a log scale) to ExpressVPN, 
NordVPN, Surfshark, Private Internet Access (PIA), and Atlas VPN ads leads to 
an increased chance of familiarity with that specific brand
by a factor of 1.41, 1.51, 2.30, 1.98, and 1.70, respectively. 
Selected models for each 
brand are shown in~\autoref{tab:vpn_brand_exposure}; brand exposure 
variables for each model were significant predictors of 
familiarity (visible in~\autoref{fig:brand_exposure}).

\begin{figure}[t]
    \centering
    \includegraphics[width=.95\linewidth]{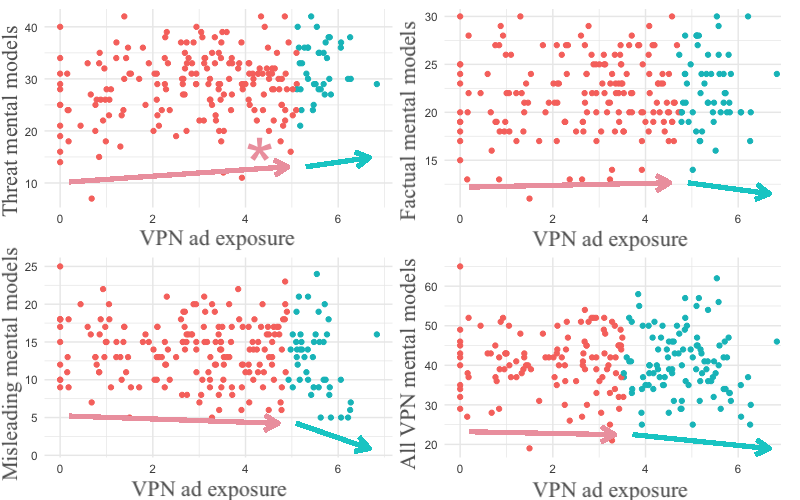}
    \caption{Two-lines analysis. Regression lines (separated by the 
    breakpoint) were color-coded and shifted for clarity.
    Left regression on \threatmm fits significantly.}
    \label{fig:two_lines}
\end{figure}

Being less tech-savvy 
is significantly associated with lower familiarity with NordVPN, 
and was selected in SurfShark and PIA models. Higher privacy sensitivity 
is significantly associated with higher familiarity with NordVPN, and appears 
in the ExpressVPN model. \new{This likely indicates that privacy sensitive or 
tech savvy people are more familiar with \sp products.} 
No other covariate is significant or 
consistent in its effect direction between models, \new{preventing 
further takeaways}. Nonetheless, 
some were selected in various individual models: 
ad interval was selected in the ExpressVPN 
model, and ``having a VPN installed'' appeared in the NordVPN 
and AtlasVPN models.

To the best of our knowledge, this is the first time security \& privacy 
awareness has been directly linked to security \& privacy 
media exposure outside of self-report measures. Our dataset 
likely captured the fundamental desired effect of ads:
establishing familiarity with a brand.

\subsection{Mental models and exposure to VPN ads}
After we establish a baseline relationship between 
exposure to VPN ads and what participants think, 
we move on to our main analysis. Following our 
plan, we first conduct two lines analysis 
for each of our four mental model variables 
with respect to exposure to VPN ads. This analysis 
does not show clear effects, the potential wear-out effects 
do not seem to lead to a significant (p$>$0.05) decline in 
mental model metrics after a breakpoint. 
Conversely, the estimated regression lines for \threatshort and 
\misleadingmm (\autoref{fig:two_lines}) 
suggests a weak but consistent trend across 
the VPN ad exposure range. Thus, we modify our analysis plan: we 
fit one (instead of two) linear regression model for the entire 
VPN ad exposure range for each mental model variable, similar to the 
analysis conducted with brand familiarity. We follow the same 
model selection procedure as described before (\autoref{sec:analysis}).

Our final results are given in~\autoref{tab:mental_models}. 
The selected model for \threatmm shows a significant and strong 
relationship between exposure to VPN ads and \threatmm: leading to an 
estimated 5.7 point change in threat mental models (in an effective 
range of 35) between the participants with the least and most exposure. 
This effect exists even though the selected model controls for 
privacy sensitivity (IUIPC-8), technical expertise (tech advice 
frequency), and VPN ownership history.

No other variable was a significant predictor of 
any mental model variable; however, we note the following 
observations from selected covariates: higher 
factual VPN mental models might be associated with higher privacy 
sensitivity (p=$0.08$), technical expertise (p=$0.12$), 
and owning a VPN (p=$0.07$); higher misleading VPN mental models 
might be associated with lower VPN ad exposure (p=$0.08$) 
and not owning a VPN (p=$0.19$). Mean VPN ad interval was 
selected in the Misleading mental models model and IUIPC-8 was 
selected in all VPN mental models model but neither 
was significant.

To understand users' mental models of VPNs in greater detail, each
participant was asked to justify one answer selected at random from 
the Likert-type questions about VPN mental models \new{(see~\autoref{tab:statements})}.
This qualitative analysis supplements our quantitative
analysis by highlighting aspects of user's VPN mental models that 
closed-ended questions might not capture.
Two researchers qualitatively coded these 217 free 
responses into one of five categories that captured how users justified their
Likert-type answer: 
some VPNs offer different features than others (20.7\%), 
VPNs improve things in the average case (13.4\%), 
VPNs are incapable of performing a task (25.8\%), 
the task is an inherent function of a VPN (32.3\%), 
or participant's free response conflicted with or failed to justify their 
Likert-type response (7.8\%). 12.4\% (10/81) of participants who were asked about misleading statements said 
that was an inherent feature of VPNs, though most of these
respondents expressed having a lack of knowledge on the subject and
erred towards VPNs being capable of accomplishing tasks.
15.6\% (10/32) of respondents who were asked about features often bundled with
VPNs (such as password leak notifications) thought that those features were
not add-ons, but core VPN capabilities.
The opposite held true as well; 9.6\% (10/104) of respondents 
who were asked about factual claims \emph{underestimated} 
the capabilities of VPNs.
These included misconceptions of encryption, such as ``a VPN does not
change or encrypt traffic, only reroutes it'' and ``I believe it encrypts
at their data centers, not my device.''
Finally, 10 participants explicitly cited advertisements as a source
of knowledge, suggesting \vpnad impact, albeit small, on mental models.

\begin{figure}
    \centering
    \includegraphics[width=.85\linewidth]{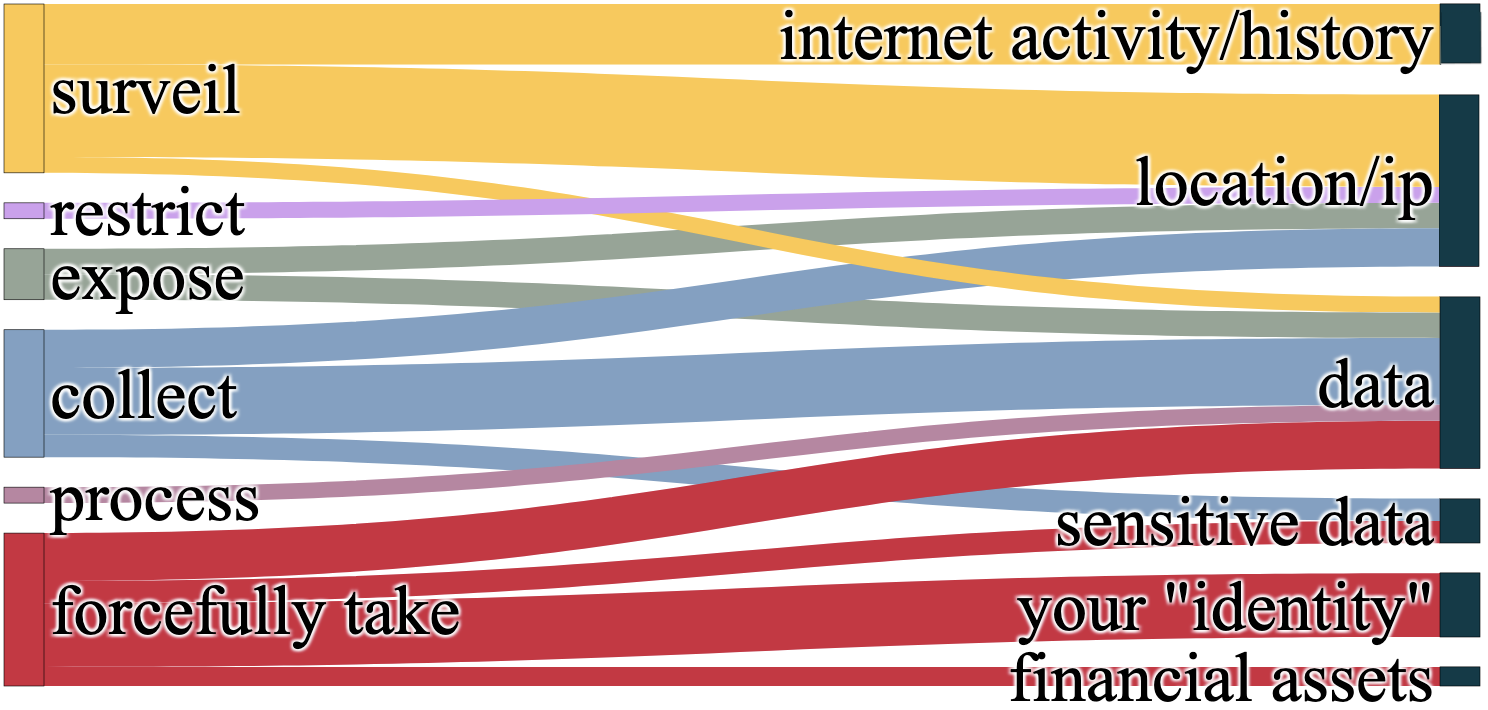}
    \caption{Attack-asset pairs participants said VPNs would protect against. 
        Band width corresponds to number of responses. 
    Pairs with fewer than five responses not shown.}
    \label{fig:threat_models}
\end{figure}

\subsection{What do VPNs protect against?}
To better understand threat protection perceptions, we asked participants to 
list (open-ended) the two most severe threats they believe VPNs protect against.
This question allows us to compare threat models of our participants
to those advertised in \vpnads~\cite{akgul2022vpn}. 
We analyzed responses 
using the same methods 
we outlined previously~\cite{akgul2022vpn},
determining attacks, adversaries, and assets.
We find that while there is significant overlap in 
all three categories, notable differences do exist. 

\paragraph{Adversaries}  
Similar to \vpnads, participants most frequently named 
``hackers'' (8.2\% of participants), 
vague adversaries (e.g., ``bad guys'', 8.0\% of participants), and 
vague companies (e.g., ``companies'', 4.0\%) as the adversaries VPNs protect 
against. Surprisingly, our participants gave much more emphasis on 
malware compared to \vpnads (5.2\% of participants vs. 
none noted~\cite{akgul2022vpn}).
However, unlike \vpnads, participants rarely named 
governments (1.7\%) or ISPs (2.4\%) as adversaries. 
Further, 66.8\% did not name any adversary. 
\new{The focus on malware might be explained 
by end users commonly using antivirus solutions~\cite{ion2015hack}.}

\paragraph{Attacks and assets}
In contrast to adversaries, 
participants were much more likely to list assets protected (64.7\%) and 
threats mitigated by VPNs (72.1\%), often in the same response. 
~\autoref{fig:threat_models} depicts this relationship.

Participants listed many fewer unique assets, with different frequency, than ads did. 
Location/IP protection 
was most commonly noted (18.5\%), often associated with surveillance, unwanted 
exposure and collection. ``Data'' (14.7\%) and ``sensitive data'' (5.8\%) 
was often collected, or forcefully taken (e.g., ``stolen'').
Many participants were convinced that VPNs would protect their 
identities (5.6\%), especially against identity theft (4.6\%).
In contrast, 
our prior work
reports that internet activity 
is the most commonly mentioned asset in \vpnads, followed by 
``data'', nebulous security and privacy concepts (e.g., ``privacy,'' 
``safety''), and ``yourself.'' Identity theft does not appear to be a common 
threat in \vpnads.

\paragraph{The (non)impact of \vpnad exposure}
Though we found similarities between the threat models of our participants
and those advertised in \vpnads, we hypothesize that this effect should 
be stronger among those who were exposed to more \vpnads. To test this, 
we first obtained the popularity ranking of each asset, attack, and adversary
among \vpnads from 
our previous work~\cite{akgul2022vpn}.
We then ran a series of regression analyses between exposure to \vpnads and 
the ranking of assets, attacks, and adversaries reported 
by our participants within the \vpnad popularity list. We expected statements 
from participants with more exposure to rank higher. However, 
we find no statistically significant relationship, 
implying that even if this effect exists, it is likely weak.

\section{Concluding discussion}
\label{sec:discussion}
We explored the impact of \vpnads on users' mental models.
We measured mental models through specially developed questionnaires 
and measured exposure to ads by analyzing users' YouTube histories 
with a purpose-built BERT-based ad detector.
Our results show that \yt users are extensively exposed to \vpnads, 
which we found to strongly correlate with brand familiarity 
and increased belief in \note{hyperbolic} threats advertised.
However, we find no significant relation between ad 
exposure and specific mental models of VPNs, \new{this includes 
misleading mental models (\autoref{tab:statements}).}
We discuss the implications of our findings below.

\paragraph{\Sp media \new{exposure correlates with} mental models: a new form of evidence}
Our analysis \new{suggests} that \vpnads likely impact users, partially confirming 
our
prior speculation~\cite{akgul2022vpn}. Participants 
who saw more ads for any of the five specific VPN brands also were 
significantly more likely to be more familiar with the respective brand. 
Further, we find that increased belief in \note{(hyperbolic)} threat 
statements made in \vpnads is linked with increased exposure to \vpnads. 
To the best of our knowledge, this is the first time exposure to 
\sp media was measured through non-self-report means and explicitly 
linked to mental models, confirming findings from previously exclusively 
self-report studies~\cite{fulton2019media, baig2021media, 
redmiles2016learn, routi2017posture}.

\paragraph{\new{Correlation with} threats but not VPN mental models}
Unlike \threatmm, our analysis does not provide conclusive evidence 
linking exposure to \vpnads with increased belief in specific VPN mental models 
expressed in \vpnads.
This observation \new{may be} due to emotional appeals (such as those in threat 
statements~\cite{akgul2022vpn}) in ads having a stronger effect on 
consumers than technical appeals~\cite{obermiller2005ad}.
Our qualitative results suggest that the technical 
information in \vpnads might affect users (they answered VPN-related questions 
while referencing ads as their source), 
but isn't potent enough to produce statistically significant results.

\paragraph{\new{What do VPN ads do}}
\new{We argue that the relationship between brand familiarity 
and exposure is due to ads increasing familiarity with a brand.
This phenomenon is well studied in prior work~\cite{schmidt2015advertising}.}
\new{This relationship isn't as straightforward with mental models. Ads 
could be contributing to the development of these models, 
or reinforcing them for users who already hold them.
Regardless, our results show that while technical details might not
be memorable, threat models are, perhaps as a result
of seeing ads or perhaps because advertisers (or YouTube content 
delivery algorithms) target users
with these models. Awareness of threats can benefit users via increased 
vigilance, but \note{hyperbole} can create excessive fear; further research is 
needed to determine what the right balance is.}

\paragraph{\new{Misleading mental models}}
\new{We observe that a large number of participants
(see~\autoref{fig:two_lines}, bottom left) believe 
in misleading mental models about VPNs. This observation 
has been echoed in prior work~\cite{story2021awareness}}
\new{Though our work does not find a direct correlation 
between these the misleading mental models and exposure 
to VPN ads that mention these models~\cite{akgul2022vpn},
this does not mean they are harmless. 
These ads might still mislead mental models, but in less obvious ways.
For instance, users might increase confidence in their 
misinformed mental models through exposure to such ads.}

\new{Misinformed mental models may result in poor 
decision making when adopting VPNs. 
For instance, users might purchase a VPN when they don't 
need one (e.g., using TOR over VPNs~\cite{fassl2023vpnfolklore}); 
or worse, might think they are protected against certain 
threats when they aren't (e.g., thinking VPNs prevent credit 
card misuse~\cite{story2021awareness}).}

\paragraph{\new{Consumer education}} \new{Misinformed mental models
could perhaps be mitigated through consumer education. Our work, along with 
prior work~\cite{bai2020improving}, hints that 
technical details might not be memorable in short interventions. 
Further, it indicates that interstitial media might not be 
the right platform to communicate nuanced security mental models. Similar 
to the public health approach to medicine, early intervention before 
undesirable outcomes occur might be necessary 
(e.g.,~\cite{reynolds2007effects, campbell2014early}).
As previous research has 
noted~\cite{blinder2024evaluating,akgul2021evaluating}, 
perhaps early educational curricula should incorporate 
appropriate use cases for \sp tools.}

\paragraph{Recommendations for future work} Though our study provides concrete evidence of the 
relationship between \sp media and mental models, it is not without limitations. 
We do not explore a multitude of variables that could affect advertising 
effectiveness~\cite{schmidt2015advertising}. Further, our study is limited to \yt and \vpnads. Though we 
expect similar results from other \sp media, this is not a foregone conclusion.
By exploring these factors, researchers can clarify the exact limits of 
\sp media influence.

\section*{Acknowledgments}
We would like to thank our participants for taking the time and sharing 
their data. Our reviewers and our 
shepherd provided 
feedback that improved this work, 
we thank them.
Thanks to Mitchell Smith for helping with early 
versions of the VPN ad classifier and dataset.
This research was supported in part by NSF grants CNS-1943240 and CNS-2323193, as well as a seed grant from the Maryland Cybersecurity Center.

\bibliographystyle{IEEEtran}
\bibliography{bib_vpn.bib}

\newpage
\appendix 
\appendixpage
\label{sec:appendix}

\section{Screener survey}
\label{app:screener}
\begin{itemize}
    \item The purpose of this survey is to determine eligibility for the rest of the study.
    \item \textit{[A consent form was shown to the participants. They were asked to consent to continue with the survey.]}
    
   \item Which of the following social media sites do you use on a regular basis (at least once a month)? Choose any that apply.
      
   \hspace{.1cm} Facebook -- YouTube -- Twitter -- TikTok -- Instagram

   \item What operating system is on your primary computer?
     
   \hspace{.0cm} Windows -- MacOS -- I don't have a computer -- Other

\item What operating system is on your primary smartphone?

\hspace{.1cm} iOS -- Android -- I don't have a smartphone -- Other
    
    \item Please estimate how many YouTube videos you watch. \textit{When answering please consider the past one year.}
\begin{itemize}
       \item More than 10 videos everyday
       \item 10 videos a day - 5 videos a day
       \item 5 videos a day - 1 video a day
       \item 1 video a day - 1 video every 3 days
       \item 1 video every 3 days - 1 video a week
       \item I don't watch YouTube
    \end{itemize}

    \item Please estimate how often \textbf{you} are signed in to your primary account when watching YouTube. \textit{When answering please consider the past one year.}
    \begin{itemize}
       \item I'm always signed in
       \item I'm frequently signed in
       \item I'm sometimes signed in
       \item I'm rarely signed in
       \item I'm never signed in
    \end{itemize}

    \item Please estimate how often \textbf{other people} (friends, family, kids etc.) watch YouTube videos while signed in to your primary account \textbf{without you also watching}. \textit{When answering please consider the past one year.}

    \hspace{.1cm} Always -- Frequently -- Sometimes -- Rarely -- Never

    \item Has your YouTube watch history been on for your primary account for the past year? \textit{You can check by going to https://www.youtube.com/feed/history and verifying that your history is visible.}

    \hspace{.1cm} Yes -- No

    \item For how long have you had this account?
    \begin{itemize}
       \item Less than a year
       \item One year - two years
       \item Two years - three years
       \item More than three years
    \end{itemize}

    \item This is a two stage study. If found eligible, we will require you to share data with researchers for research purposes. 
    
    We built measures to ensure the \textbf{data will only be accessible to the researchers} on this project. \textbf{The data will be permanently deleted} after researchers finalize relevant reports. 
    
    \textit{Please select all that you are willing to share. You may select none if you do not wish to continue beyond this eligibility survey.}
    \begin{itemize}
       \item {[Required for full study]} Your YouTube watch history and subscriptions.
       \item {[Required for full study]} The list of programs currently installed on your primary computer.
       \item {[Required for full study]} Your history of downloading apps on your primary Android or Apple devices.
    \end{itemize}

\end{itemize}

\section{Tutorial stage}
\label{app:tutorials}
\begin{itemize}
   \item \textit{[A consent form was shown to the participants. They were asked to consent to continue with the survey.]}

    \item This study consists of two stages. You will complete stage 1 today. At this stage (the first stage) you will only be compensated for completing the questionnaires, and following data export tutorials (\$2.65-\$3.65 in total).

    \textbf{The full compensation for the data will be sent only if you successfully complete both stages of the study.} You can expect to have earned \textbf{\$9.45-\$10.45} when you complete the entire study.

    \item Please rate your agreement or disagreement with the following statements. There are no right or wrong answers, we are only interested in what you think. \textit{[Asked on a seven-point Likert scale: Strongly Agree - Agree - Somewhat Agree - Neutral - Somewhat Disagree - Disagree - Strongly Disagree]}

    \begin{itemize}
       \item Consumer online privacy is really a matter of consumers' right to exercise control and autonomy over decisions about how their information is collected, used, and shared.
       \item Consumer control of personal information lies at the heart of consumer privacy.
       \item I believe that online privacy is invaded when control is lost or unwillingly reduced as a result of a marketing transaction.
       \item Companies seeking information online should disclose the way the data are collected, processed, and used.
       \item A good customer online privacy policy should have a clear and conspicuous disclosure.
       \item It usually bothers me when online companies ask me for personal information. 
       \item When online companies ask me for personal information, I sometimes think twice before providing it. 
       \item I'm concerned that online companies are collecting too much personal information about me. 
    \end{itemize}

    \item Have you ever used a Virtual Private Network (VPN) before? \textit{Please select all that apply.}
    \begin{itemize}
       \item Yes, I have used a VPN for personal purposes.
       \item Yes, I have used a VPN for work purposes.
       \item No, I have never used a VPN.
    \end{itemize}

    \item \textit{[The following two questions were displayed if the user selected yes to the previous question.]}
    
    \item Did you have to pay to use a VPN for personal purposes (not for work)?

    \hspace{.1cm} Yes -- No

    \item Please write the name(s) of the VPN(s) you have used.

    \item How familiar are you with the following brands? {[\textit{Asked on a 1-7 scale with 1 being not at all familiar and 7 being very familiar. The icons for each brand was also displayed.}]} 

    \hspace{.1cm} ExpresssVPN -- NordVPN -- Surf Shark -- PIA -- Atlas VPN

    \item In your opinion, what are the two most severe threats a VPN can protect you against? Please list the threats from most to least severe. \textit{There are no right or wrong answers, we are only interested in what you think.} {[\textit{Two responses were required for this question; however, three text boxes were given.}]} 

    \item In your opinion, what are the two most important benefits from most to least important. \textit{There are no right or wrong answers, we are only interested in what you think.} {[\textit{Two text boxes were given for this question.}]} 
    
    \item Thanks for completing the study so far. We will now continue on with data export tutorials.

\end{itemize}

[\textit{Particiapnts were shown a combination of the 
following series of tutorials depending on their devices. Tutorials were combined 
where appropriate.}]
\begin{itemize}
   \item \textit{Exporting your YouTube history and subscriptions}
   \item \textit{Exporting your history of downloading apps on your primary Android devices}
   \item \textit{Exporting your history of downloading apps on your primary Apple devices}
   \item \textit{Exporting the list of programs currently installed on your primary Windows computer}
   \item \textit{Exporting the list of programs currently installed on your primary MacOS computer}
\end{itemize}

\section{Final stage}
\label{app:final_stage}
\begin{itemize}
    \item This is the final stage of this study. Thank you for taking the time and sticking with the study so far!

    This stage will consist of uploading the remaining data (which you had requested in the previous stage) and a questionnaire on your perceptions of the internet. You will only be able to complete this stage (and therefore eligible for pay) if we receive all the data we initially requested in the first stage. We estimate that this stage will take about six minutes to complete. You will be compensated \textbf{\$1.20} for the survey and \textbf{\$1.75} per data source you provide. You can expect to earn \textbf{\$6.45-\$8.20} in this stage.

    Click next to proceed.

    \item \textit{[One of the following two were shown based on whether we had received all data requested so far.]}
    \item Please upload the zip file(s) containing the data here [interface shown in~\autoref{fig:upload_portal}]. If the data is split between multiple zip files, please upload all. To limit the data you are sharing with us, even if the zip file(s) contain additional data, this upload portal will only upload data that is relevant to our study.
    
    The portal shows you which data sources we have not received yet (yellow boxes), once all data is uploaded ("Data received: 100\%") you can click next.

    \item Thanks for successfully uploading all of the data in the previous stage. Next, you will be asked to complete the final questionnaire. No technical knowledge is required, we are only interested in your opinions. Please click next to proceed.

    \item Have you heard of VPNs (Virtual Private Networks) before?

    \hspace{.1cm} Yes -- No -- Unsure

    \item \textit{[The following two questions were only displayed if the respondent selected yes in the previous question.]}

    \item Where have you heard of VPNs from before?

    \item Have you ever seen advertisements for VPNs embedded in online videos you watched?
    
    \hspace{.1cm} Yes -- No -- Unsure

    \item \textit{[The following question was only displayed if the respondent selected yes in the previous question.]}

    \item Where have you seen advertisements for VPNs in videos you watched? Please select all that apply.

    \hspace{.1cm} YouTube -- Twitter -- TikTok -- Instagram -- Facebook -- Television -- other

    \item \textit{[The following question was only displayed if the respondent selected YouTube in the previous question.]}

    \item Please estimate how many VPN advertisements you have seen embedded in YouTube videos in the past year.

    \begin{figure}
      \centering
      \includegraphics[width=.9\linewidth]{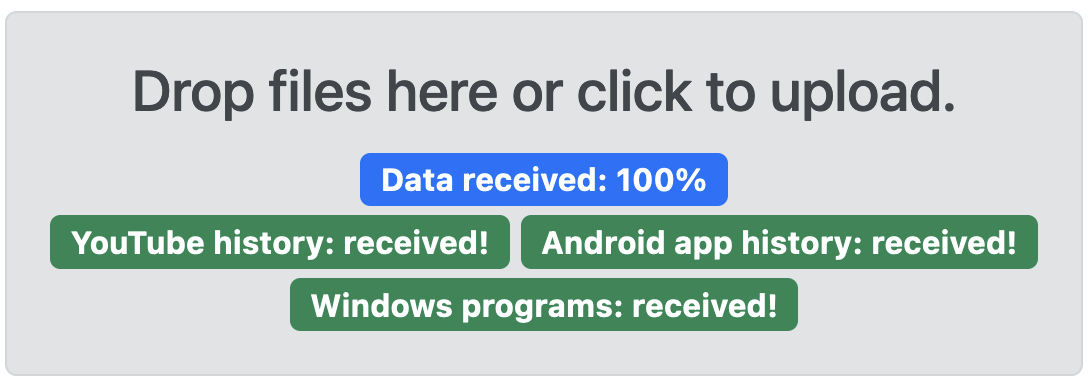}
      \caption{The upload portal interface. Indicates all required data sources were received.}
      \label{fig:upload_portal}
    \end{figure}

    \item \textit{[The following question was only displayed if the respondent selected responded affirmatively that they had seen advertisements for VPNs embedded in online videos they watched.]}

    \item Please list the three most important features being advertised about VPNs embedded in YouTube videos you have watched. {[\textit{Two responses were required for this question; however, three text boxes were given.}]}

    \item Please indicate how much you agree or disagree with the following statements. \textit{There are no right of wrong answers, we are only interested in what you think.} \textit{[Asked on a seven-point Likert scale: Strongly Agree - Agree - Somewhat Agree - Neutral - Somewhat Disagree - Disagree - Strongly Disagree]}

    \hspace{.1cm} \textit{[see threat mental models in~\autoref{tab:statements}]}

    \item In your opinion, which of the following statements is true for consumer VPNs? \textit{There are no absolute right of wrong answers, we are only interested in what you think.} \textit{[Asked on a five-point Likert-type scale: True for all VPNs - True for almost all VPNs - True for some VPNs - True for almost no VPNs - True for no VPN]}
    
    \hspace{.1cm} \textit{[see VPN mental models in~\autoref{tab:statements}]}

    \item \textit{[The following free-response question was based on a random statement chosen form the previous question. Participants were then asked to explain their response to this statement. An example question would look like:]} 

    Please explain why you said "True for almost all VPNs" for "I can watch other countries streaming libraries (e.g., Netflix, Hulu) if I use a VPN."

    \item Please indicate your age. If you'd prefer not to answer, you can skip this question.

    \item What gender do you best identify with?
    \begin{itemize}
       \item Man
       \item Woman
       \item Prefer to self-describe
       \item Prefer not to say
    \end{itemize}

    \item Which of the following best decribes your race? Select all that apply.
    \begin{itemize}
       \item White
       \item Black or African American
       \item American Indian or Alaskan Native
       \item Asian
       \item Hispanic or Latino
       \item Native Hawaiian or Pasific Islander
       \item Other
       \item Prefer not to say
    \end{itemize}

    \item Please specify the highest degree of level of school you have completed or currently attending.
    \begin{itemize}
        \item No high school degree
        \item High school graduate, diploma or the equivalent (for example, GED)
        \item Some college credit, no degree
        \item Trade, technical, vocational training
        \item Associate's degree
        \item Bachelor’s degree
        \item Master’s degree
        \item Professional degree
        \item Doctorate degree
        \item Other
        \item Prefer not to say
    \end{itemize}

    \item What is your current employment status? Select all that apply.
    \begin{itemize}
        \item Employed Full-Time
        \item Employed Part-Time
        \item Self-employed
        \item Unemployed
        \item Student
        \item Home-maker
        \item Retired
        \item Other
        \item Prefer not to say
    \end{itemize}

    \item What is your annual household income?
    \begin{itemize}
        \item Up to \$25,000
        \item \$25,000 to \$49,999
        \item \$50,000 to \$74,999
        \item \$75,000 to \$99,999
        \item \$100,000 or more
        \item Prefer not to say
    \end{itemize}

    \item How frequently do you give computer or technology advice (e.g., to friends, family, or colleagues)?

      \hspace{.1cm} Always -- Often -- Sometimes -- Rarely -- Never

\end{itemize}

\end{document}